
\magnification=\magstep1
\font\mfonta=cmbx10 scaled \magstep1
\font\mfontb=cmr10  scaled \magstep0
\def\m@th{\mathsurround=0pt }
\def\ref#1{[\hbox{#1}]}
\def\unt#1{$\setbox0=\hbox{#1} \dp0=0pt \m@th \underline{\box0}$}
 \def\o17{$^{17}O$} \def\f18{$^{18}F$}
\outer\def\beginpar#1\par{\vskip0pt plus .3\vsize\penalty-250
 \vskip0pt plus -.3\vsize\bigskip\bigskip\bigskip\vskip\parskip
 \message{#1}\leftline{\bf#1}\nobreak\bigskip\noindent}
\footline={\hfill}

\vskip 1true cm
\centerline{\mfonta Correlations in Nuclei:}
\bigskip
\centerline{{\mfonta Self-Consistent Treatment and the BAGEL Approach}
\footnote{$^{\star}$}{Supported by the Commission
of the European Communities }}
\bigskip
\bigskip\bigskip\bigskip
\centerline{{\mfontb H. M\"uther and  L.D. Skouras}
\footnote{$^\dagger$}{Permanent address: Institute of Nuclear Physics,
N.R.C.P.S.
{\it Demokritos,} Aghia Pa\-ras\-kevi GR 15310, Greece}
 }
\bigskip
\centerline{{\it Institut f\"ur Theoretische Physik der Universit\"at
 T\"ubingen}}
\centerline{{\it D-7400 T\"ubingen, Federal Republic of Germany}}
\bigskip\bigskip\bigskip\bigskip\bigskip\bigskip
\midinsert\narrower
\noindent {\bf Abstract}:
\noindent
An approach is presented which allows a self-consistent description of
the fragmentation of single-particle strength for nucleons in finite
nuclei employing the Greens function formalism. The self-energy to be
considered in the Dyson equation for the single-particle Greens
function contains all terms of first (Hartree-Fock) and second order in
the residual interaction. It is demonstrated that the fragmentation of
the single-particle strength originating from the terms of second order
can efficiently be described in terms of the so-called BAGEL
approximation. Employing this approximation the self-energy can be
evaluated in a self-consistent way, i.e. the correlations contained in
the Greens function are taken into account for the evaluation of the
self-energy. As an example this scheme is applied to the nucleus $^{16}O$,
using a realistic nucleon nucleon interaction. The effects of
the correlations on the occupation probabilities and the binding energy
are evaluated.

\endinsert
\vfill\eject
\baselineskip=16pt
\headline={\hss -- \folio\ --\hss}
\noindent
Most of the microscopic nuclear structure studies are based on the
independent particle model (IPM) describing the nucleus as a system of
nucleons moving without any correlations in a mean field (Hartree-Fock
potential) generated by the interaction with all other nucleons. A lot
of effort has been made, both form the experimental as well as the
theoretical side, to explore the borderlines of this simple shell-model
and search for clear fingerprints of correlations beyond the IPM.
Nucleon knock-out experiments with electrons or hadrons have been
performed\ref{1-4} to determine absolute spectroscopic factors and
study deviations of occupation probabilities from the predictions of
the IPM.

The depletion of occupation numbers for hole-states, states which are
completely occupied in the IPM, partly originates in the strong
short-range components of a realistic nucleon-nucleon (NN) interaction.
The effects of these short-range correlations have mainly been studied
for nuclear matter\ref{5-8} assuming that the main results can be
transferred to finite nuclei. Here we want to focus the attention on
long-range correlations. This implies that one has to consider a
finite nucleus since such correlations are sensitive to the low-energy
excitation spectrum, reflecting the shell-structure of finite nuclei.
The effects of short-range correlations are taken into account by evaluating
an effective interaction  appropriate for the model-space under
consideration. As described in ref.\ref{9} this is done by solving the
Bethe-Goldstone equation for a realistic OBE potential\ref{10} assuming
a Pauli operator appropriate for this model-space.

As a starting point of the discussion let us consider the definition of
the irreducible self-energy for the nucleons taking into account the terms of
first and second order in the effective interaction ${\cal V}_{eff}$.
The first order contribution can be written
$$\Sigma^{(1)}_{\alpha\beta} = \sum_{\gamma \gamma '}
 \int {d \omega_1 \over 2 \pi i}
< \alpha \gamma \vert {\cal V}_{eff} \vert \beta \gamma ' >
G_{\gamma\gamma '}
(\omega_{1}) \; , \eqno(1)$$
while the terms of second order are defined by
$$\eqalignno{\Sigma^{(2)}_{\alpha\beta}(\omega ) = & {1 \over 2}  \sum_{\gamma
\delta \mu \gamma ' \delta '\mu '} \int {d \omega_1 \over 2 \pi i}
\int {d \omega_2 \over 2 \pi i}
\cr
& \times < \alpha \mu \vert {\cal V}_{eff} \vert \gamma '\delta '> < \gamma
\delta \vert {\cal V}_{eff} \vert \beta \mu '> G_{\gamma\gamma '}(\omega -
\omega_1 + \omega_2 ) G_{\delta \delta '} (\omega_1 ) G_{\mu \mu '}
(\omega_2)\; . &(2)
\cr}$$
The summations on single-particle quantum numbers $\gamma,
\delta \dots $ are restricted to orbits which define the model-space.
In our test calculation on $^{16}O$ discussed below, single-particle
states up to the 1p0f shell are taken into account. In equations (1)
and (2)
$G_{\alpha\beta}(\omega )$ refers to the Greens function which is
obtained from the Dyson equation
$$G_{\alpha \beta } (\omega ) = \delta_{\alpha \beta}
g_{\alpha} (\omega ) + \sum_{\gamma } g_{\alpha } (\omega )
\Sigma^{(2)}_{\alpha
\gamma } (\omega ) G_{\gamma \beta } (\omega ) \eqno(3)$$
and $g_{\alpha}$ is the Greens function determined for the self-energy
contribution of first order in the interaction $\Sigma^{(1)}$. It is
obvious that equations (1) to (3) have to be solved in a
self-consistent way. As a first step of  an iteration scheme to
reach this self-consistency we consider the Hartree-Fock (HF) approximation
for $g_{\alpha}$
$$ g_{\alpha} (\omega ) =  \left[ {\Theta (\epsilon^{HF}_{\alpha} -
\epsilon_{F}) \over \omega - \epsilon^{HF}_\alpha + i\eta } + {\Theta
(\epsilon_{F} - \epsilon^{HF}_\alpha )
\over \omega - \epsilon^{HF}_\alpha -i\eta }\right] \; . \eqno(4)$$
It is diagonal in the basis of HF single-particle states and defined in
terms of the single-particle energies $\epsilon^{HF}_\alpha$ for states
$\alpha$ above and below the Fermi energy $\epsilon_{F}$.
In the first iteration
step also the Greens functions $G$ in eqs.(1) and (2) are replaced by
the corresponding HF Greens functions. On this level the self-energy
terms $\Sigma^{(1)}$ and $\Sigma^{(2)}$ are represented by the Goldstone
diagrams of figure 1a) and 1b) plus 1c), respectively. Inserting this
approximation for the self-energy into the Dyson equation one can
obtain the solution by solving the following\footnote{$^{\ddag}$}
{In order to keep the notation simple we ignore here and in the following
presentation the non-diagonal terms in the self-energy.} set of
equations (see ref.\ref{9,11,12})
$$\pmatrix{\epsilon_{\alpha}^{HF} & a_1 & \ldots & a_K & b_1 & \ldots & b_L
\cr a_1 & E_1 & & &  & & \cr \vdots & & \ddots & & &  & \cr
a_K &  & & E_K & & &  \cr b_1 & &  & & \tilde E_1 & & \cr \vdots & & & &
& \ddots & \cr b_L &  & \ldots &  & & \ldots & \tilde E_L \cr }
\pmatrix{ X_0^n \cr X_1^n \cr \vdots \cr X_K^n \cr Y_1^n \cr \vdots \cr Y_L^n
\cr }
= \omega_n
\pmatrix{ X_0^n \cr X_1^n \cr \vdots \cr X_K^n \cr Y_1^n \cr \vdots \cr Y_L^n
\cr } \; , \eqno(5)$$
where all elements of the matrix that are not indicated, are zero,
$$ E_j = \epsilon_{p_1}^{HF} + \epsilon_{p_2}^{HF} - \epsilon_{h}^{HF}
\; , \eqno(6) $$
is the energy of a 2 particle 1 hole (2p1h) configuration $j$ and
$$a_{j} = < p_1 p_2 \vert {\cal V} \vert \alpha h > \; , \eqno(7) $$
the coupling of the single-particle state $\alpha$ to the 2p1h
configuration $j$. The normalization of the eigenvectors in eq.(5) to
unity gives the correct normalization of the spectroscopic factors
$(X_{0}^n)^2$. In a similar way the quantities $\tilde E_{k}$ and
$b_{k}$ refer to 2 hole 1 particle (2h1p) configurations. The Greens
function $G_{\alpha}$ is then given in the Lehmann representation as
$$G_{\alpha}(\omega ) = \sum_{n} \left( X_{0}^n \right) ^2
 \left[ {\Theta (\omega_{n} - \epsilon_{F}) \over \omega - \omega_{n}
+ i\eta } + {\Theta (\epsilon_{F} - \omega_{n} )
\over \omega - \omega_{n} -i\eta }\right] \; . \eqno(8)$$
Due to the coupling of the single-particle state $\alpha$ to the 2p1h
and 2h1p configurations the single-particle strength is distributed to
various discrete states at energies $\omega_{n}$. In general one will
observe for single-particle orbits $\alpha$, which in the IPM are below
the Fermi level, an occupation probability
$$n_{\alpha} = \sum_{n} \Theta (\epsilon_{F} - \omega_{n})
\left( X_{0}^n \right) ^2 \; , \eqno(9)$$
which is different from one. Similarly, one obtains a non-zero
occupation probability for single-particle states above the Fermi level.

In a second
iteration step one may try to employ the Greens function of eq.(8) in
the definition of the self-energy. This is no problem for the
$\Sigma^{(1)}$. Rather than the conventional BHF single-particle
energies we will now obtain
$$\epsilon_{\alpha}^{RHF} = <\alpha \vert t \vert \alpha> + \sum_{\beta}
< \alpha \beta \vert {\cal V}_{eff} \vert \alpha \beta  > n_{\beta} \;
. \eqno(10) $$
Translated into the language of diagrams: The modification of the
Greens function due to the second order terms in the self-energy yields
a depletion of the occupation of the hole states, as represented in
lowest order by diagram 1d), and an occupation larger than zero for
"particle" states as expressed by 1e). This correction is similar to
the Renormalized Brueckner Hartree Fock approach ( therefore we use
the label RHF in eq.(10)) accounting for depletion of holes and
occupation of particle states\ref{8,13}.

It requires much more effort to use the improved Greens function also
in the definition of the second order self-energy (see as examples the
diagrams of figures 1f) and 1g)). Again the solution of the Dyson
equation (3) can be transformed into a problem of a matrix
diagonalisation. However, replacing the single-pole approximation of the
HF Greens function of eq.(4) by the Greens function of eq.(8) leads to
an explosion of basis configurations in the matrix equation which will
replace eq.(5) in this second iteration step. Each combination of
energies $\omega_{n}$ for the various orbits $\alpha$ yields a new
configuration to be considered in this matrix. This increase of the
number of configurations reflects the fact that in this second
iteration step beside 2p1h configurations also 3p2h, 4p3h etc
configurations are taken into account as indicated by the diagrams 1f)
and 1g).

Van Neck et.al. have proposed a scheme to represent the distribution of
the single-particle strength in terms of strength selected in energy
bins\ref{14}.  Here we would like to  follow a different route, based
on the assumption that for practical purposes the Greens function of
eq.(8) is well approximated by a small number of
poles, which are representative for the spectral distribution at low
energy. To introduce the method we recall that eq.(5) may be rewritten
in a basis
in which the matrix is tri-diagonal
$$\pmatrix{\Omega_{K}& \alpha_{K} & 0 & & & & & \cr
\alpha_{K} & \Omega_{K-1} & \alpha_{K-1} & & & & & \cr
0 & & \ddots & & & & & \cr
& & & \Omega_{1} & \alpha_{1} & & & \cr
& & & \alpha_{1} & \epsilon_{\mu}^{HF} & \beta_{1} &0 & \cr
& & & 0 & \beta_{1} & \tilde \Omega_{1} & \beta_{2} & & \cr
& & & & & \ddots & & \cr
& & & & & & \beta_{L} & \tilde \Omega_{L}\cr}
\pmatrix{\xi_{K}^n \cr \xi_{K-1}^n \cr \vdots \cr \xi_{1}^n \cr
X_{0}^n \cr \zeta_{1}^n \cr \vdots \cr \zeta_{L}^n \cr }
=\omega_{n}
\pmatrix{\xi_{K}^n \cr \xi_{K-1}^n \cr \vdots \cr \xi_{1}^n \cr
X_{0}^n \cr \zeta_{1}^n \cr \vdots \cr \zeta_{L}^n \cr }
\; . \eqno(11) $$
This tri-diagonal form of the matrix is obtained by applying the so-called
BAGEL (BAsis GEnerated by Lanczos method\ref{9,15}) procedure to
single-particle or single-hole states $\mu$. In eqs (5) and (11)
$K$ and $L$ denote
the total numbers of 2p1h and 2h1p configurations which have the same symmetry
quantum numbers (parity, isospin and angular momentum) as the single-particle
orbit $\mu$.  In this BAGEL basis we may
truncate the matrix displayed in eq. (11). We call it the BAGEL(p,q)
approximation if, in addition to the 1-body state $\mu$, we
restrict the basis to include only the first  p combinations of 2p1h
configurations (with diagonal elements $\Omega_{1} \dots \Omega_{p}$)
together with the first q combinations of 2h1p configurations
($\tilde \Omega_{1} \dots \tilde \Omega_{q}$). If $\mu$ refers to a
hole state in the IPM we will consider p combinations of 2h1p configurations
plus q combinations of 2p1h configurations in addition to the
single-particle state $\mu$.
Diagonalization in this truncated basis yields p+q+1 eigenvalues and we
may define the Greens function in the BAGEL(p,q) approximation
by restricting the summation in eq. (8) to the
eigenvalues generated in the truncated basis.

It is obvious that BAGEL(0,0) yields the HF approximation for the
Greens function and BAGEL($K$,$L$) refers to the untruncated set of
equations (5).
The BAGEL(0,1) approach is the simplest
approximation leading for each $\mu$ to an eigenvalue below and above
$\epsilon_{F}$ which consequently yields non-trivial results for the
occupation probabilities defined in eq.(9). It is easy to verify that
the BAGEL(p,p) approximation reproduces the moments of the spectral
distribution
$$S_{m} = \sum_{n} \omega_{n}^m \left( X_{0}^n \right) ^2 \; , \eqno(12)$$
evaluated for the Greens function of the untruncated equations
from order $m = 0$ up to the
order $m = 2p+1$. Therefore this BAGEL scheme provides a set of approximations
for the Greens function in terms of p+q+1 poles, which can
systematically be improved ranging from the HF approximation (1 pole) to the
result of the untruncated eq.(5).
As an example figure 2 displays the distribution of
single-particle strength $(X_{0}^n)^2$ obtained for the 0s1/2 state in
$^{16}O$ in various BAGEL(p,q) approximations.

The key point of the procedure presented here is that a BAGEL(p,q)
approximation is inserted in the calculation of the self-energy (2).
This self-energy then contains a much larger number of poles than  when
the single-pole approximation(4) is inserted in (2). The Dyson equation
may still bee written in a matrix form (5). The dimension of the matrix
is much larger. Therefore an exact solution is out of the question. The
BAGEL truncation scheme can still be applied. Truncating again with the
same values for p and q, an iterative scheme is obtained, which yields
a self-consistent solution of eqs.(1-3).
As an example let us consider the
single-particle state $\alpha$ = 0s1/2 for $^{16}O$ in the model-space
discussed above. In the first iteration step, i.e. approximating $G$ by
the HF or BAGEL(0,0) approach the dimension is $K+L+1$ = 82. If now for
the second iteration step we employ the BAGEL(p,q) approximation for
(p,q)=(0,1),(1,1),(2,2) or (6,6) the corresponding dimension is raised
to 596, 2178, 9638 or 160558, respectively. But even for a matrix of
dimension 160558 it is not very difficult to determine the BAGEL(6,6)
approximation to be used in the next iteration step and continue until
convergence is achieved.

In order to demonstrate the sensitivity of the results to the choice of
p and q in this scheme, table 1 exhibits some typical results obtained
in a self-consistent calculation employing various BAGEL(p,q)
approximations. For the hole states we present the occupation
probabilities $n_{\alpha}$ as defined in eq.(9) and the mean value for
the hole-energy:
$$\epsilon_{\alpha}=
\omega_{\alpha}^< = { 1 \over n_{\alpha}} \sum_{n} \Theta (\epsilon_{F}
- \omega_{n}) \omega_{n}
\left( X_{0}^n \right) ^2 \; , \eqno(13)$$
while for selected particle states we show the corresponding mean value for
the particle energies. Furthermore also the calculated binding energy
per nucleon
$$ {E \over A} = {\sum_{\alpha} (2j_{\alpha}+1) \sum_{n} { 1 \over 2}
\left[ <\alpha
\vert t\vert \alpha> + \omega_{n}\right] \Theta (\epsilon_{F} - \omega_{n})
\left( X_{0}^n \right) ^2 \over \sum_{\alpha} (2j_{\alpha}+1)
n_{\alpha}} \; , \eqno(14)$$
is given.

The correlation effects included in this scheme produce
remarkable differences compared to the HF (or IPM) approach. The
occupation probabilities for the hole states are reduced by about 3
percent and the single-particle energies for these states are more
attractive by around 1 to 2 MeV. The total binding energy is increased
by around 1 MeV per nucleon. A fast convergence is obtained comparing
the different BAGEL approximations. Already the simplest approach
beyond the IPM, the BAGEL(0,1) scheme, yields a result very close to
the most sophisticated one (p=q=6), which we have been studying.

In order to demonstrate the importance of self-consistency, the sixth
column of table 1 shows results obtained in the BAGEL(6,6) approach
after the first iteration step, i.e considering the HF approximation
for the Greens function in calculating the self-energy. This implies
that only terms as displayed in figures 1a) - 1c) are taken into
account. One finds that this first iteration tends to overestimate the
importance of correlations.

The last column of table 1 shows the single-particle energies obtained
in the RHF approximation according to eq.(10). At this level diagrams
of figure 1a), 1d) and 1e) are taken into account. One finds that the
inclusion of the occupation probabilities according to 1d) and 1e)
increases the single-particle energies of all states as compared to HF.
The additional inclusion of diagrams displayed in figures 1b), 1c), 1f)
and 1g) decreases the energies for the hole states but increases the
energies of particle states. Therefore one enhances the gap between
mean values for particle and hole energies, as defined in eq.(13),
at the Fermi level.
This should not be confused with the fact \ref{16} that correlations decrease
the gap in the excitation spectrum, as those energies are related to
the solutions of the Dyson equation that are closest to the Fermi
energy.

The results of these studies demonstrate that the BAGEL scheme
introduced here yields a very powerful tool for the representation of
the single-particle Greens function, which can be used to solve the
problem of calculating self-energy and Greens function in a
self-consistent way. This scheme should produce reliable results also
for heavier nuclei and model spaces larger than the one considered
here. Larger model spaces and inclusion of residual interaction between
the 2p1h and 2h1p configurations tend to enhance the correlation
effects\ref{9}. A scheme very similar to this approach for finite systems
could also be developed for studies of infinite systems like nuclear
matter.

\beginsection References

\frenchspacing
\item{[1]} S. Frullani and J. Mougey, {\it Adv. in Nucl. Phys.} {\bf 14}
(1984) 1
\item{[2]} P.K.A. de Witt Huberts, {\it Nucl. Phys.} {\bf A507} (1990) 189c
\item{[3]} J.W.A. den Herder, H.P. Blok, E. Jans, P.H.M. Keizer. L. Lapikas,
E.N.M. Quint, G. van der Steenhoven, and P.K.A. de Witt Huberts, {\it Nucl.
Phys.} {\bf A490} (1988) 507
\item{[4]} P. Grabmayr, G.J. Wagner, H. Clement, and H. R\"ohm, {\it Nucl.
Phys.} {\bf A494} (1989) 244
\item{[5]} V.R. Pandharipande, C.N. Papanicolas, and J. Wambach, {\it Phys.
Rev. Lett.} {\bf 53} (1984) 1133
\item{[6]} A. Ramos, A. Polls, and W.H. Dickhoff, {\it Nucl. Phys.} {\bf
A503} (1989) 1
\item{[7]} O. Benhar, A. Fabrocini, and S. Fantoni, {\it Nucl. Phys.} {\bf
A509} (1989) 375
\item{[8]} H.S. K\"ohler, {\it Nucl. Phys.} {\bf A537} (1992) 64
\item{[9]} H. M\"uther and L.D. Skouras, {\it Nucl. Phys}, in print
\item{[10]} R. Machleidt, {\it Adv. in Nucl. Phys.} {\bf 19} (1989) 189.
\item{[11]} M.G.E. Brand, G.A. Rijsdijk, F.A. Muller, K. Allaart and W.H.
Dickhoff, {\it Nucl. Phys.} {\bf A531} (1991) 253
\item{[12]} W.H. Dickhoff and H. M\"uther, {\it Rep. Prog. Phys.} {\bf
11} (1992) 1947
\item{[13]} B.H. Brandow, {\it Rev. Mod. Phys.} {\bf 39} (1967) 771.
\item{[14]} D. Van Neck, M. Waroquier and J. Ryckebusch, {\it Nucl. Phys.}
{\bf A530} (1991) 347
\item{[15]} H. M\"uther, T. Taigel and T.T.S. Kuo, {\it Nucl. Phys.} {\bf
A482} (1988) 601
\item{[16]} C. Mahaux, P.F. Bortignon, R.A. Broglia and C.H. Dasso,
{\it Phys. Rep.} {\bf 120} (1985) 1
\vfil\eject
\beginsection Table 1:

Results for the single-particle energies (see eq.(13)) and the
occupation probabilities for selected states in $^{16}O$ as obtained in
self-consistent BAGEL(p,q) calculations. Also listed are the energies
per nucleon calculated according to eq.(14). The last 2 columns contain
the results obtained in BAGEL(6,6) after the first iteration and the
RHF single-particle energies according to eq.(10). All energies are listed
in MeV.
\vskip 2 true cm
{\offinterlineskip \tabskip=0pt
\halign{ \strut \vrule #& \quad # \quad & \hfil # \quad & \vrule # &
\quad \hfil # & \quad \hfil #  &  \quad \hfil # & \quad \hfil #
& \quad \hfil # \quad &\vrule # & \quad \hfil # & \quad \hfil #
 \quad & \vrule # \cr
\noalign{\hrule}
&&&&&&&&&&&&\cr
&\multispan 2 \hfil $\alpha$ \hfil && HF & B(0,1) & B(1,1) &
B(3,3) & B(6,6) && 1 Iter & RHF &\cr
&&&&&&&&&&&&\cr
\noalign{\hrule}
&&&&&&&&&&&&\cr
&$s_{1/2}$& $\epsilon$ && -44.15 & -45.14 & -45.19 & -45.10 & -45.09
&& -45.98 & -43.33 &\cr
&& $n_{\alpha}$&& 1.000 & 0.982 & 0.981 & 0.978 & 0.978 && 0.979 &&\cr
&&&&&&&&&&&&\cr
&$p_{3/2}$& $\epsilon$ && -20.82 & -22.04 & -22.06 & -22.18 & -22.17 &&
-22.71 & -20.37 &\cr
&& $n_{\alpha}$&& 1.000 & 0.977 & 0.976 & 0.966 & 0.965 && 0.965 &&\cr
&&&&&&&&&&&&\cr
&$p_{1/2}$& $\epsilon$ && -17.34 & -19.14 & -19.19 & -19.43 & -19.43
&& -19.87 & -17.00 &\cr
&& $n_{\alpha}$&& 1.000 & 0.968 & 0.967 & 0.952 & 0.951 && 0.951 &&\cr
&&&&&&&&&&&&\cr
&$d_{5/2}$& $\epsilon$ && -1.77 & -1.01 & -0.86 & -0.84 & -0.83 &&
-0.95 & -1.64 &\cr
&& $n_{\alpha}$&& 0.000 & 0.010 & 0.011 & 0.016 & 0.017 && 0.018 &&\cr
&&&&&&&&&&&&\cr
\noalign{\hrule}
&&&&&&&&&&&&\cr
&E/A&&& -5.015 & -6.054 & -6.063 & -6.030 & -6.018 && -6.343 &&\cr
&&&&&&&&&&&&\cr
\noalign{\hrule} }}

\hfil \vfil \eject
\beginsection Figure Captions

\bigskip\noindent
\item{Figure 1:} Contributions to the self-energy $\Sigma^{(1)}$ (a, d,
e) and $\Sigma^{(2)}$ (b, c, f, g). While the diagrams a) - c) show the
contributions obtained in the first iteration step (replacing the
Greens functions in eq.(1) and (2) by the HF approximation), the
diagrams d) - g) show examples of contributions generated through the
self-consistent scheme.

\bigskip\noindent
\item{Figure 2:} Distribution of the single-particle strength ($\left(
X_{0}^n \right) ^2$ for the various poles $\omega_{n}$ in the Greens
function) as obtained in different BAGEL(p,q) approximations for the
$0s_{1/2}$ state in $^{16}O$. Note the logarithmic scale and the fact
that the strength obtained in BAGEL(1,1) at positive energies is hidden
by the corresponding result obtained in BAGEL(0,1) as both are
essentially identical.

\bye